\documentclass[12pt,preprint]{aastex}

\shorttitle{A Massive Evolved Galaxy at $z$ = 1.26}
\shortauthors{Matsuoka et al.}

\begin{document}


\title{Optical to Near-IR Spectrum of a Massive Evolved Galaxy \\ at $z$ = 1.26}


\author{Y. Matsuoka\altaffilmark{1,2}, B. A. Peterson\altaffilmark{3}, S. Oyabu\altaffilmark{4},
K. Kawara\altaffilmark{1}, N. Asami\altaffilmark{1}, \\ H. Sameshima\altaffilmark{1}, N. Ienaka\altaffilmark{1}, 
T. Nagayama\altaffilmark{5}, and M. Tamura\altaffilmark{6}}


\altaffiltext{1}{Institute of Astronomy, The University of Tokyo, 2-21-1, Osawa,  Mitaka, Tokyo 181-0015, Japan; 
  matsuoka@ioa.s.u-tokyo.ac.jp.}
\altaffiltext{2}{Research Fellow of the Japan Society for the Promotion of Science.}
\altaffiltext{3}{Mount Stromlo Observatory, Research School of Astronomy and Astrophysics, Australian National University, 
  Weston Creek P.O., ACT 2611, Australia.}
\altaffiltext{4}{Institute of Space and Astronautical Science, Japan Aerospace Exploration Agency, Sagamihara, Kanagawa
    229-8510, Japan.}
\altaffiltext{5}{Department of Astronomy, Kyoto University, Sakyo-ku, Kyoto 606-8502, Japan.}
\altaffiltext{6}{National Astronomical Observatory of Japan, 2-21-1, Osawa, Mitaka, Tokyo 181-8588, Japan.}


\begin{abstract}
We present the optical to near-infrared (IR) spectrum of the 
galaxy TSPS J1329$-$0957, a red and bright member of the class of 
extremely red objects (EROs) at $z$ = 1.26. This galaxy was found in the 
course of the Tokyo-Stromlo Photometry Survey (TSPS) which we 
are conducting in the southern sky. 
The spectroscopic observations were carried out with the Gemini 
Multi-Object Spectrograph (GMOS) and the Gemini Near Infra-Red 
Spectrograph (GNIRS) mounted on the Gemini-South telescope.
The wide wavelength coverage of 0.6 -- 2.3 $\mu$m provides useful
clues as to the nature of EROs while most published spectra 
are limited to a narrower spectral range which is dictated by 
the need for efficient redshift determination in a large 
survey.  We compare our spectrum with several optical composite 
spectra obtained in recent large surveys, and with stellar 
population synthesis models. 
The effectiveness of using near-IR broad-band data, instead of
the spectral data, in deriving the galaxy properties are also investigated.
We find that TSPS J1329$-$0957 formed 
when the universe was 2 -- 3 Gyr old, and subsequently evolved 
passively to become one of the most massive galaxies found in 
the $z$ = 1 -- 2 universe. Its early type and estimated stellar mass 
of $M_{*} = 10^{11.5} M_{\odot}$ clearly point to this galaxy being a 
direct ancestor of the brightest elliptical and spheroidal 
galaxies in the local universe.
\end{abstract}





\section{Introduction\label{sec:intro}}

The formation and evolution of massive (stellar mass $M_* > 10^{11} M_{\odot}$) galaxies represent
one of the key issues in understanding the whole picture of mass assembly in the universe.
This class of objects is believed to correspond to the local early-type galaxies, i.e., ellipticals and spheroidals,
constituting the brightest end of the luminosity function.
They are predicted to form gradually with time through merging of smaller subsets in the hierarchical 
structure formation scenario based on the $\Lambda$ cold dark matter (CDM) model \citep[e.g.,][]{white78, cole94}.
While the $\Lambda$CDM model can successfully reproduce the large scale structure of galaxies,
difficulties appear
when it deals with  smaller systems such as individual galaxies.
Although the hierarchical scenario predicts the time of appearance of the most massive galaxies at $z \sim 1$,
there is growing evidence that fully assembled galaxies exist at $z >1$ \citep[e.g.,][]{dunlop96, spinrad97, 
daddi04, fontana04, glazebrook04, cimatti04, saracco05}.
They are largely passively evolving, early-type galaxies containing old stellar populations 
\citep[e.g.,][]{yan03, yan04, mccarthy04, mignoli05, longhetti05}.
The alternative scenario of mass assembly is represented by the monolithic model, in which galaxies are
formed in a single collapse and star burst, and subsequently evolve passively \citep[e.g.,][]{larson74, 
tinsley77, bruzual80}.
In this framework even the most massive galaxies could exist in very early universe.
While both the hierarchical and monolithic scenarios are consistent with the observations in the local universe,
their differences should become evident at $z > 1$.
Recently \citet{conselice07} showed that the observed mass growth of massive galaxies from $z = 2$ to
$z = 0$ is significantly different from the prediction of the Millennium Simulation \citep{lemson06, delucia06} based 
on the $\Lambda$CDM model.

Thus motivated by their
unique importance, the massive galaxies beyond redshift $z = 1$ 
have been a subject of
extensive studies over the last decade.
Several large survey programs have been completed or are ongoing, 
such as
the Gemini Deep Deep Survey \citep[GDDS;][]{abraham04}, the K20 Survey \citep{cimatti02}, and others 
\citep[e.g.,][]{doherty05, saracco05, roche06}.
The $z > 1$ massive galaxies exhibit very red optical to near-IR colors, either due to evolved stellar populations, or 
severe reddening in case of dusty starbursts, and they 
form a class of
extremely red objects \citep[EROs;][]{elston88}.
This ERO nature is commonly utilized to pick up the objects from multi-color photometry catalogs for 
subsequent
spectroscopy, usually carried out in the red-visible (0.6 -- 0.9 $\mu$m) light, to obtain the objects' redshifts 
\citep[e.g.,][]{mccarthy04b}.

While the recent studies have been achieving great successes in revealing the mass assembly history out to the redshift 
$z >$ 2, and placing significant constraints on the galaxy formation scenarios, one should be aware of the fact that most
of these works are based on optical spectroscopic observations supplemented with near-IR broad-band photometry, usually fitted
with the spectral energy distribution (SED) models constructed from the local galaxy populations.
It is based on the implicit assumption that the optical 
to IR SED, and its evolution with age, of the $z > 1$ galaxies are well represented by those of local counterparts.
However, such an assumption has not yet been observationally confirmed due to the lack of the data with sufficiently large (i.e., from
optical to near- or mid-IR) spectral coverage.
The same shortage of the data has prevented the empirical-template method to be used in the analysis of the observed
(spectral or broad-band) SEDs in wide wavelength ranges.
Although recent works based on the large surveys have constructed some composite spectra by stacking all the obtained data,
such composites have an apparent disadvantage in that different parts of the spectra are represented by systematically different
galaxy populations, e.g., blue part by the high-redshift objects and red part by the low-redshift objects.
Thus, the accumulation of the observations and analysis of wide spectral regions for individual galaxies are definitely needed at this 
stage in order to make comprehensive studies of the massive galaxy population existing at $z > 1$.

In this paper we report the first example of that kind of data, a red-visible to near-IR (0.6 -- 2.3 $\mu$m) spectrum of the massive evolved galaxy
TSPS J1329$-$0957 at $z$ = 1.26.
It has been discovered in the course of the Tokyo-Stromlo Photometry Survey (TSPS) which we are conducting in
the high galactic-latitude sky observed from the southern hemisphere.
Introduction of the TSPS, target selection, and the observations and reductions of the optical and near-IR spectra are detailed in \S \ref{sec:obs}.
We present in \S \ref{sec:char} the reduced spectrum which is compared to optical composite spectra generated in the past surveys and to
stellar population synthesis models.
Our comparisons 
reveal the basic properties of TSPS J1329$-$0957 as a representative of  passively-evolving galaxies found in the $z$ = 1 -- 2 universe.
The effectiveness of using near-IR broad-band data, instead of the spectral data, in deriving the galaxy properties are also investigated.
Finally a brief summary appears in \S \ref{sec:summary}.
Throughout this paper we use the Vega magnitude system and a standard cosmology of $H_{\rm 0}$ = 70 km s$^{-1}$ Mpc$^{-1}$, $\Omega_{\rm M}$ = 0.3, and 
$\Omega_{\Lambda}$ = 0.7.

\section{Observations\label{sec:obs}}


\subsection{Target Selection: the Tokyo-Stromlo Photometry Survey}

The Tokyo-Stromlo Photometry Survey (TSPS) is a multi-color ($IZJHK_s$) photometry survey of the high galactic-latitude ($| b | > 30^{\circ}$) sky
observed from the southern hemisphere.
The main targets of the survey are bright extra-galactic sources.
It has been developed by the collaboration of  teams in The University of Tokyo and in the Research School of Astronomy and Astrophysics, 
The Australian National University (ANU).
The survey is divided into two parts; the $I$- and $Z$-band survey observation and the near-IR (usually $JHK_s$-band) follow-up photometry.

The $IZ$-band survey started in 2003.
More than 1000 deg$^2$ has been observed to date in photometric conditions using the Wide Field Imager (WFI) on the ANU 40-inch Telescope and 
the drift-scan camera on the United Kingdom Schmidt Telescope (UKST), both located at the ANU's  Siding Spring Observatory (SSO).
Observations with the WFI are carried out under the typical seeing condition of around 2\arcsec\ while the pixel scale of the camera, with the
configuration of 2 $\times$ 2 pixel binning, is 0.76\arcsec/pixel.
The standard 5$\sigma$ limiting magnitudes are $I_{\rm Vega}$ = 21.4 and $Z_{\rm Vega}$ = 20.5.
The performance and data products of the drift-scan camera on the UKST are currently under examination.
To date over 150 nights have been devoted to the $IZ$-band survey.

Parts of the  surveyed fields are subsequently observed in the near-IR ($JHK_s$) bands 
for the purpose of discovering high-redshift ($z > 5.7$) quasars.
We aim at exploring various aspects of the early universe such as the growth of super massive black holes related to 
galaxy formation \citep[e.g.,][]{magorrian98, haring04}, the re-ionization of intergalactic medium by means of the Gunn-Peterson trough 
\citep[e.g.,][]{gunn65, loeb01, fan06}, and star formation activity traced by chemical-abundance analysis \citep[e.g.,][]{hamann93, kawara96, yoshii98, 
matsuoka05, matsuoka07, matsuoka08, tsuzuki06}.
Two telescopes/instruments are mainly used for the near-IR observations; the Cryogenic Array Spectrometer/Imager (CASPIR) on the ANU 2.3-m Advanced 
Technology Telescope at the SSO, and the SIRIUS camera \citep{nagashima99, nagayama03} on the Infrared Survey Facility (IRSF) 1.4m telescope located 
at Sutherland, South African Astronomical Observatory.
The sky conditions of the CASPIR observations are similar to those of the WFI observations while the pixel scale of the camera is 0.5\arcsec/pixel.
On the other hand, typical seeing of the SIRIUS observations, with the pixel scale of 0.453\arcsec/pixel, is around 1\arcsec.
5$\sigma$ limiting magnitudes range from ($J$, $H$, $K_s$)$_{\rm Vega}$ = (20.0, 19.4, 18.1) to (20.7, 20.1, 18.8) depending on the apparent 
brightness of the high-redshift quasar candidates in the surveyed field\footnote{
Magnitudes in the near-IR bands are expressed in 2MASS photometric system throughout this paper. 
See Nakajima, Y. et al., in preparation, for the conversion of magnitudes and colors between IRSF/SIRIUS and 2MASS systems.}.
All the near-IR observations are carried out in photometric conditions.
We have devoted another 150 nights, so far, to the near-IR observations.
The main features and results of the TSPS project will be presented in detail in forthcoming papers (e.g., Asami, N., et al., in preparation).

\begin{figure}
\epsscale{0.80}
\plotone{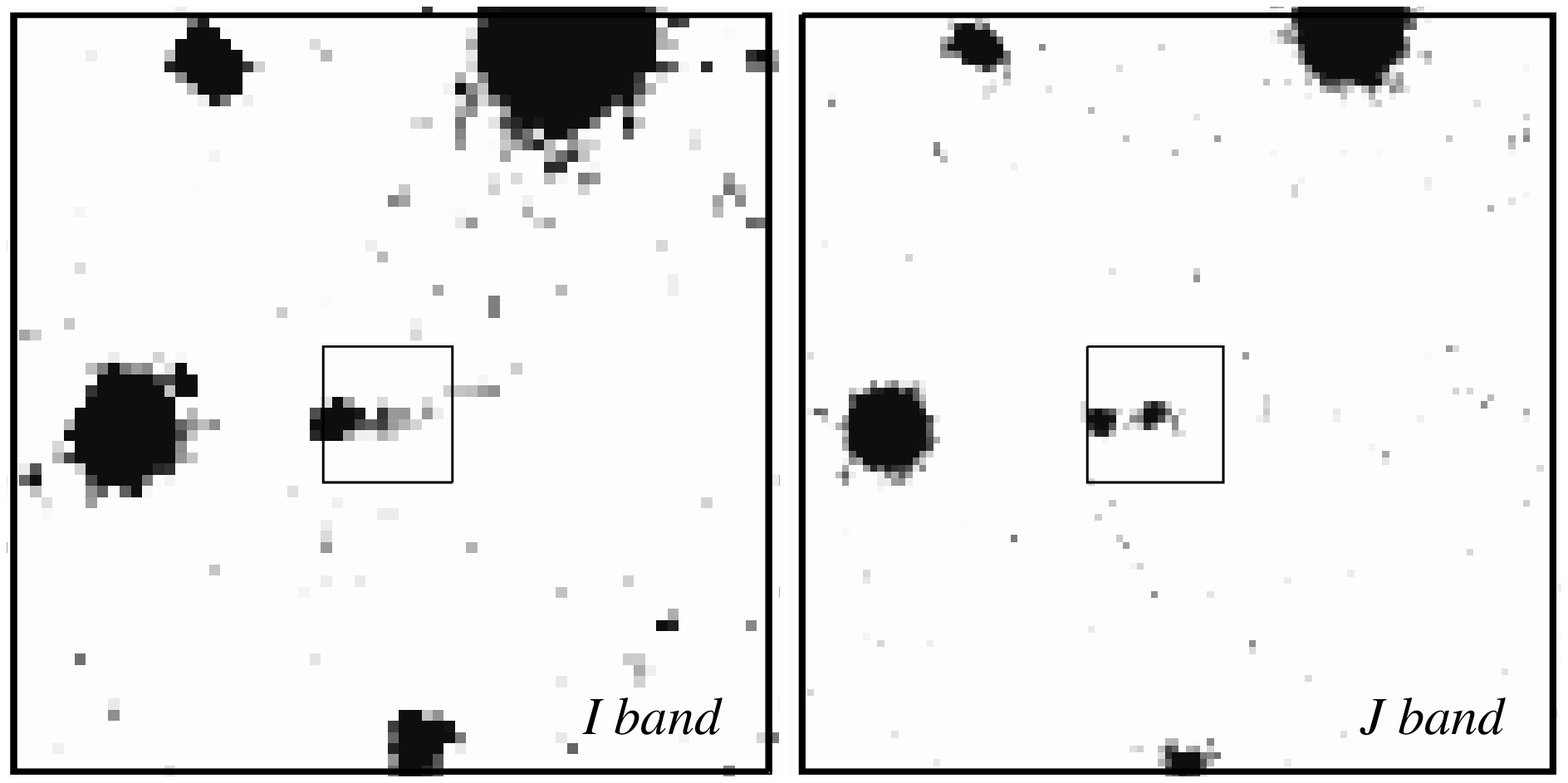}
\caption{
WFI $I$-band ({\it left}) and SIRIUS $J$-band ({\it right}) images
of TSPS J1329$-$0957 at the center of the small box.
North is up, east to the left.
The field is approximately 50\arcsec\ $\times$ 50\arcsec.\label{findchart}
}
\end{figure}

TSPS J1329$-$0957 has been discovered at R.A. 13:29:44.52, Decl. $-$09:57:29.8 (J2000.0) in a TSPS field owing to its extremely red color.
The $IZ$-band images were obtained with the WFI on 2005 February while the $JHK_s$ images were taken with the SIRIUS on 2006 March.
Its apparent magnitudes, measured within the 4.5\arcsec\ aperture, are $I$ = 21.7, $Z$ = 20.9, $J$ = 19.7, $H$ = 18.6, and $K_s$ = 17.6.
Since its color and magnitudes ($I-K_s$ = 4.1 and $K_s$ = 17.6) clearly show that TSPS J1329$-$0957 is a fairly red and bright member of EROs
\citep[$R-K > 5.3$ or $I-K > 4$; see also, e.g., Fig. 6 of][]{abraham04},
we decided to obtain its optical to near-IR spectrum in order to further investigate its nature.
The $I$- and $J$-band images of the galaxy are shown in Figure \ref{findchart}.

\subsection{Optical Spectroscopy}

The optical spectrum of TSPS J1329$-$0957 was obtained on 2007 February 18 with the Gemini Multi-Object Spectrograph
\citep[GMOS;][]{hook04} mounted on the Gemini-South telescope\footnote{
This work is based on observations obtained at the Gemini Observatory, which is operated by the
Association of Universities for Research in Astronomy, Inc., under a cooperative agreement
with the NSF on behalf of the Gemini partnership: the National Science Foundation (United
States), the Science and Technology Facilities Council (United Kingdom), the
National Research Council (Canada), CONICYT (Chile), the Australian Research Council
(Australia), CNPq (Brazil) and SECYT (Argentina).}
(Program ID: GS-2007A-Q-14).
The R400\_G5325 grating was used with the RG610\_G0331 filter, 
at two slightly different central wavelengths (791.0 nm and 794.0 nm) in
order to fill the CCD gaps.
A  slit width of  1.0\arcsec\ was used in  average seeing of 0.8\arcsec.
This configuration gives a wavelength coverage of 0.6 -- 1.0 $\mu$m and a resolution of $R \sim 1000$.
The object was dithered along the slit in the classical A-B-B-A pattern for better sky subtraction.
The total exposure time was 3.5 hours, constituting  32 400-sec exposures.

Data reduction was performed in a standard manner using the GEMINI.GMOS package within the reduction software IRAF\footnote{IRAF 
is distributed by the National Optical Astronomy Observatories, which are operated by the Association of Universities for Research
in Astronomy, Inc., under cooperative agreement with the National Science Foundation.}.
Wavelength calibration was established using  CuAr arc spectra.
The sensitivity function along the dispersion axis was obtained from the observed spectrum of the white dwarf EG 274,
taken with the same instrumental configuration as the target observations.
The calibrated spectrum of EG 274 and the atmospheric extinction data at the Cerro Tololo Inter-American 
Observatory contained in the IRAF database were used to calculate the sensitivity function.
Since the galaxy possibly exhibits an extended feature whose morphology is dependent on the observed wavelength,
though the morphology in the $IZ$-band images is not very clear due to the poor seeing ($\sim$2\arcsec) of the WFI observations
(see Fig. \ref{findchart}), aperture correction is not expected to be achieved properly by referring to the spectroscopic standard star.
We assumed the wavelength dependence of the slit loss of the form $\lambda^{-\alpha}$ where the power index $\alpha$ was determined so
that the final spectral slope is consistent with the broad-band $I-Z$ color obtained from the imaging observations ($\alpha$ = 3.5).
While the choice of function shape is rather arbitrary, it is the simplest form, and we found that, as shown below, the 
resultant optical spectrum is completely consistent with the composite spectra obtained in  past surveys.
Finally the calibration for absolute flux scale was obtained from the broad-band photometric data.

\subsection{Near-IR Spectroscopy}

The near-IR spectrum of TSPS J1329$-$0957 was obtained during 4 nights in 2007 February and March (Program ID: GS-2007A-Q-14).
The Gemini Near Infra-Red Spectrograph \citep[GNIRS;][]{elias98} mounted on the Gemini-South telescope was used in the cross-dispersed mode, 
which covers 0.9 -- 2.5 $\mu$m simultaneously.
The average seeing was 0.6\arcsec.
The adopted 31.7 l/mm grating and 1.0\arcsec\ slit with the short (0.15\arcsec/pixel) camera give a resolving power of $R \sim 500$.
Two off-source (sky) positions were observed by slightly offsetting the telescope perpendicular to the slit in order to obtain sky spectra
which were used for subtracting sky emission from the object frames.
Total exposure time on source is 2.0 hours constituting of 16 450-sec exposures.

Data reduction was carried out with the GEMINI.GNIRS package within the IRAF in a standard manner.
Distortion and wavelength calibrations were achieved with the pinhole and Ar arc spectra, respectively.
The A0V- and F0V-type stars, HIP 63109 and HIP 63151, were observed as  telluric standard stars to which blackbody spectra
with appropriate temperatures were fitted in order to calculate the sensitivity along the dispersion axis.
The additional aperture correction described in the previous section were not applied since the galaxy shows fairly concentrated shapes
with  spatial profiles similar to those of nearby stars in the $JHK_s$ bands.
The spectral slope beyond $\sim$1.2 $\mu$m agrees with the broad-band $J-H$ and $H-K$ colors within the measurement errors.
We do not use the GNIRS spectrum at wavelengths shorter than $\sim$1.2 $\mu$m since the appropriate broad-band (e.g., $Y$-band) imaging
data is not available for checking the slit loss.
As in the optical spectrum, the calibration of the  absolute flux scale was obtained from the broad-band photometric data.

\section{Results and Discussion\label{sec:char}}

\subsection{Reduced Spectrum}

\begin{figure}
\epsscale{0.80}
\plotone{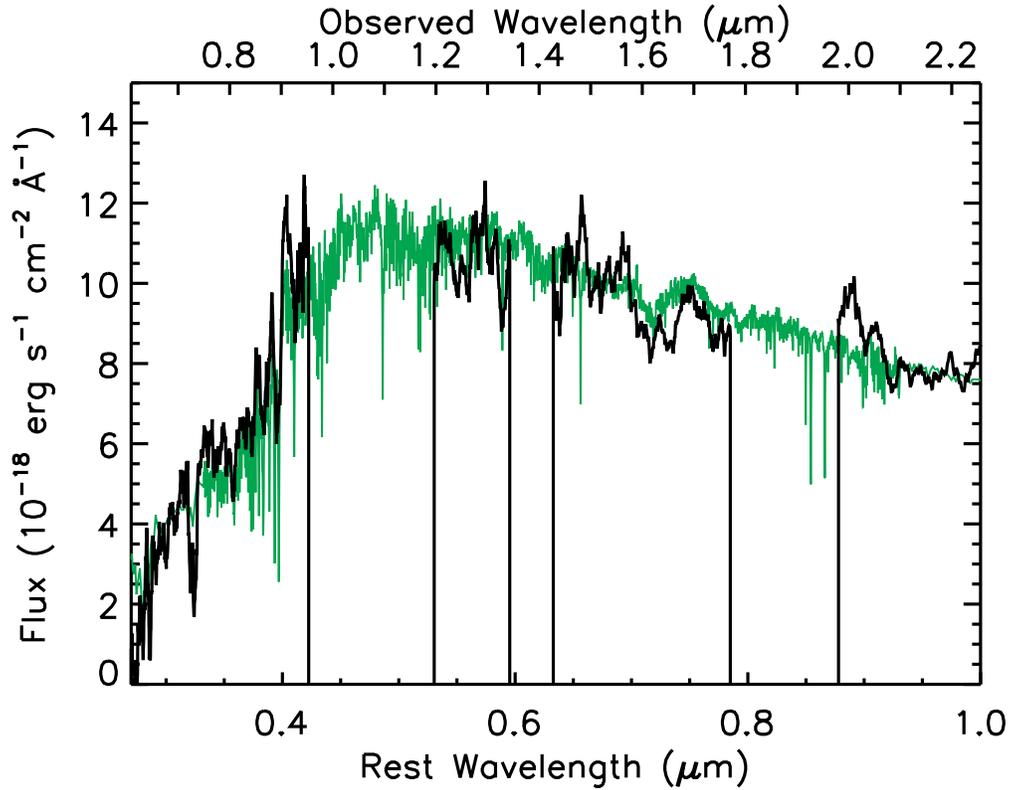}
\caption{The reduced GMOS and GNIRS spectrum of TSPS J1329$-$0957 in the rest frame ({\it black}). 
  The upper axis denotes the observed wavelength.
  Also shown is the best-fitting SPS model calculated with the \citet{bc03} code ({\it green}; see text).
  \label{full_spec}}
\end{figure}

\begin{figure}
\epsscale{0.80}
\plotone{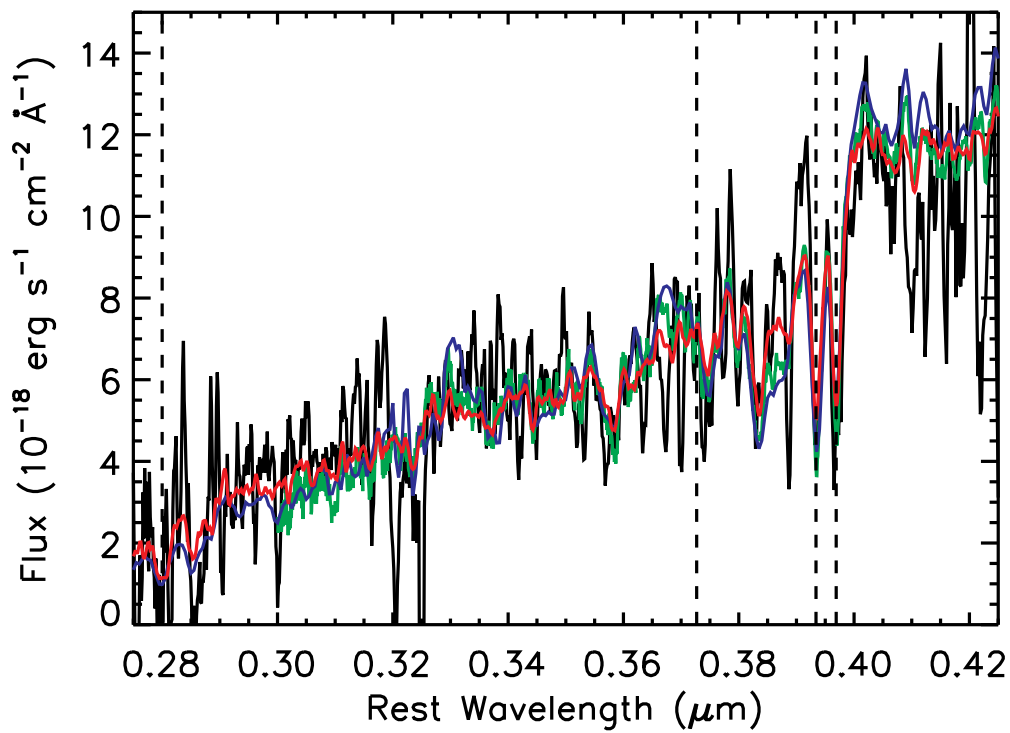}
\caption{The observed spectrum of TSPS J1329$-$0957 in the rest frame ({\it black}) compared with 
  the composite spectra generated from the K20 early-type galaxies ({\it red}), the SDSS LRGs at 
  $z$ = 0.30 -- 0.35 ({\it green}), and the local early-type galaxies ({\it blue}). 
  Dashed lines mark the expected positions of \ion{Mg}{2} $\lambda$2800, [\ion{O}{2}] $\lambda$3727, and \ion{Ca}{2} HK lines.
\label{composites}}
\end{figure}

The reduced GMOS and GNIRS spectrum is shown in Figures \ref{full_spec} and \ref{composites} ({\it black}).
The \ion{Ca}{2} HK absorption lines are clearly detected, giving the redshift of this object as $z$ = 1.26.
There are the absorption features at the expected positions of \ion{Mg}{2} $\lambda$2800 and \ion{Mg}{1} $\lambda$2852, though the data quality
is rather poor around these features since they are near the detector edge.
No other single absorption or emission lines can be identified while some broad features are in common with the composite spectra of 
past surveys and/or with the model predictions as shown below.
The detection of \ion{Ca}{2} HK lines and absence of any strong emission features such as [\ion{O}{2}] $\lambda$3727, as well as the overall SED, 
clearly indicate that TSPS J1329$-$0957 is an old, passively-evolving galaxy whose light is dominated by evolved stellar population.
Considering its redshift and $K_s$-band brightness ($K_s$ = 17.6 mag), this ERO is likely to be among the most massive 
galaxies with a stellar mass M$_{*}$ exceeding 10$^{11.5}$ M$_{\odot}$.
Hereafter we reveal some basic properties of the galaxy through comparisons with the optical composite spectra obtained in recent large
surveys and with the stellar population synthesis (SPS) models.

\subsection{Comparison with the Optical Composite Spectra\label{sec:optcomp}}

In Figure \ref{composites} we compare the optical spectrum of TSPS J1329$-$0957 with the composite spectra 
generated from the K20 early-type galaxies \citep[{\it red};][]{mignoli05}, the Sloan Digital Sky Survey (SDSS) Luminous Red Galaxies 
(LRGs) at $z$ = 0.30 -- 0.35 \citep[{\it green};][]{eisenstein03}, and the local early-type galaxies ({\it blue}) compiled by \citet{mannucci01}.
Each spectrum has been scaled in accordance with the flux at the wavelength interval 0.34 -- 0.36 $\mu$m and is expressed in the rest frame.
The overall feature of the spectra blueward of $\sim$4000 \AA\ are remarkably similar in spite of the systematic difference in the redshifts of galaxies 
constituting the composites, i.e., mean redshift $<z>$ = 0.75 for the K20 galaxies, $z$ = 0.30 -- 0.35 for the SDSS LRGs, and 
$z$ = 0 for the local early types, while $z$ = 1.26 for TSPS J1329$-$0957.
It clearly demonstrates the early-type nature of TSPS J1329$-$0957, containing little amount of dust responsible for the reddening of the visible light.

\begin{figure}
\epsscale{0.70}
\plotone{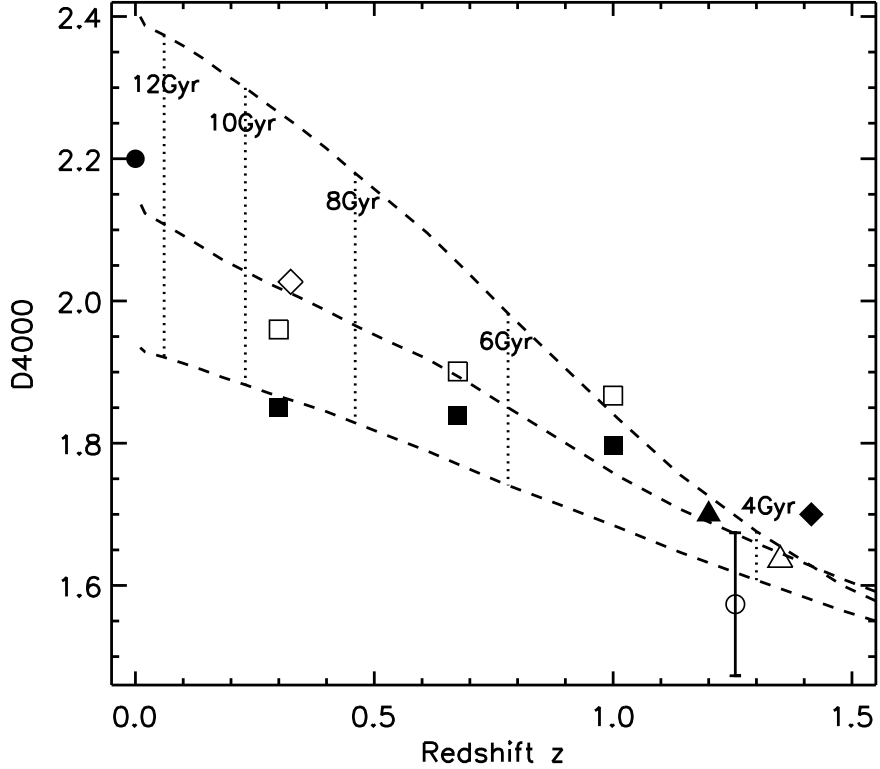}
\caption{The D4000 index, adopting the definition of \citet{bruzual83}, of various composite spectra and TSPS J1329$-$0954
  plotted as a function of redshift. The open circle with the error bar represent TSPS J1329$-$0954 while the other symbols 
  represent the composite spectra of the early-type galaxies
  ({\it filled circle}: the local galaxies, {\it open diamond}: the SDSS LRGs, {\it open squares}: the K20 bright subsamples, {\it filled squares}: 
  the K20 faint subsamples \citep[see][]{mignoli05}, {\it filled triangle}: the LCIRS galaxies,
  {\it open triangle}: the GDDS galaxies at $z$ = 1.3 -- 1.4, and {\it filled diamond}: the MUNICS ``young'' subsamples).
  See text for a full description of the references.
  Three dashed lines show the predictions of the SPS models by the \citet{bc03} code with the metallicity $Z$ = $Z_{\odot}$, 0.4 $Z_{\odot}$, 
  and 0.2 $Z_{\odot}$ from top to bottom, assuming the exponentially-declining star formation with the e-folding time $\tau$ = 1.0 Gyr and a formation
  epoch (the epoch of the onset of star formation) at redshift $z_{\rm f} = 7$.
  Dotted lines represent the redshifts corresponding to the galaxy ages of 4, 6, 8, 10, and 12 Gyr.\label{d4000}}
\end{figure}

On the other hand, a small but clear difference is seen in the 4000 \AA\ break.
We show the D4000 index, adopting the definition of \citet{bruzual83}, of various composite spectra and TSPS J1329$-$0957 in Figure \ref{d4000}.
The plotted composites are those described above plus those generated from the GDDS early-type galaxies at $z$ = 1.3 -- 1.4, the
Las Campanas Infrared Survey (LCIRS) pure early-type galaxies \citep{doherty05}, and the Munich Near-IR Cluster Survey (MUNICS) ``young'' 
early-type galaxy population defined by \citet{longhetti05}.
Also shown, are the predictions of the SPS models by the \citet{bc03} code, assuming an  exponentially-declining
star formation history with the e-folding time $\tau$ = 1.0 Gyr, the metallicity $Z$ = $Z_{\odot}$, 0.4 $Z_{\odot}$ and 0.2 $Z_{\odot}$, and 
the epoch of the onset of star formation at redshift $z_{\rm f}$ = 7.
The figure clearly shows the increase of the observed D4000 index from $z$ = 1.5 toward the local universe, which coincides with the
simple galaxy aging represented by the SPS models.
The index of TSPS J1329$-$0957 is also on this trend and is consistent with those of the GDDS galaxies and the MUNICS young subsamples within
the measurement error.
We infer the age of TSPS J1329$-$0957 to be a few Gyr.

\subsection{Comparison with the SPS Models\label{sec:sedfit}}

\subsubsection{Basic Properties of TSPS J1329$-$0957}

Here we compare the overall SED of TSPS J1329$-$0957 with the predictions of  SPS models by the \citet{bc03} code.
In the model calculations, star formation is assumed to start at $t$ Gyr before the galaxy is observed and subsequently reduce its rate 
exponentially with the e-folding time $\tau$ Gyr.
Dust content is parametrized in terms of the color excess $E_{\rm B-V}$ with the extinction law of \citet{calzetti00}.
These parameters are varied over the plausible ranges, i.e., $t$ = 0.01 -- 10.0 Gyr, $\tau$ = 0.01 -- 10.0 Gyr, and $E_{\rm B-V}$ = 
0.0 -- 1.0 mag, and the best-fitting model is searched for by the least ${\chi}^2$ method.
The total stellar mass of the galaxy is determined through normalization of the model SED.
We start with the \citet{salpeter55} initial mass function (IMF) and the solar metallicity ($Z$ = $Z_{\odot}$).
Although other sets of calculations adopting the \citet{fitzpatrick99} dust extinction raw or the \citet{chabrier03} IMF were also performed, 
they provide neither better fits nor significantly different best-fitting results.

We show the best-fitting model with the observed spectrum in Figure \ref{full_spec} ({\it green}).
The (observed-frame) optical and near-IR parts of the spectrum make similar contributions to the ${\chi}^2$ where it is minimum.
As seen in the figure, the overall agreement between the observation and the model is fairly good over the whole wavelength range.
The parameter values comprising the best-fitting model are summarized in the first row of Table \ref{sps_params}.
They are: $t$ = 2.4$^{+0.2}_{-0.5}$ Gyr, $\tau$ = 0.55$^{+0.10}_{-0.15}$ Gyr, $E_{\rm B-V}$ = 0.06$^{+0.07}_{-0.03}$ mag, and
log $M_{*}$ [$M_{\odot}$] = $11.49^{+0.01}_{-0.04}$ where the quoted errors denote 90 \% confidence.
The derived age (2 -- 3 Gyr) and dust content ($E_{\rm B-V} \la$ 0.1 mag) are quite consistent with the estimates from D4000 index and
from the comparison with the composite spectra of the early-type galaxies as described in the previous section.
We have also performed the calculations assuming the sub-/super-solar metallicity ($Z$ = 0.4 $Z_{\odot}$ and 2.5 $Z_{\odot}$) and found that 
the best-fit age of the galaxy is $\sim$1.5-Gyr older/younger than the original ($Z$ = $Z_{\odot}$) case, which is the consequence of the well-known 
age-metallicity degeneracy.
However, no better fits than the $Z$ = $Z_{\odot}$ case are achieved in these calculations.

Thus we conclude that TSPS J1329$-$0957 is very likely to be formed 2 -- 3 Gyr before it was observed, and subsequently evolved passively 
to become one of the most massive objects found in the $z$ = 1 - 2 universe.
Its formation epoch corresponds to the redshift $z_{\rm f}$ = 2.6 when the universe is about 2.5 Gyr old.
It is similar to the median formation redshift of the GDDS evolved galaxies studied by \citet{mccarthy04} ($<z_{\rm f}>$ = 2.4)
and of the MUNICS young subsamples studied by \citet{longhetti05} ($<z_{\rm f}>$ = 2.3).
The early-type nature and the estimated stellar mass of $M_{*} = 10^{11.5} M_{\odot}$ clearly indicate that the galaxy is a direct ancestor of the 
brightest elliptical and spheroidal galaxies present in the local universe.

\subsubsection{Effectiveness of Near-IR Broad-Band Data}

Most of the previous studies have utilized optical spectra supplemented with near-IR broad-band photometry in deriving the physical properties of 
EROs by the SED-model fitting.
Thus it is useful to compare the fitting results for such data sets with the above results for the whole (optical and near-IR) spectral data.
We first consider the optical spectrum with the $K$ ($K_s$)-band photometry, which is the most common data set used in the ERO studies.
The $K_s$-band magnitude was calculated from the GNIRS spectrum and the SIRIUS filter transmission curve.
We fitted the SPS-model SEDs to this data set in the same way as described above, and deduced the best-fitting parameters.
The results are summarized in the second row of Table \ref{sps_params}; the data set yields smaller $t$,  $\tau$ and $E_{\rm B-V}$ than the case of the
whole spectral data.
The shorter duration of star formation makes the galaxy SED (at a fixed age) redder while the younger age and less dust content make it bluer, 
thus the overall SED is similar to the case of the whole spectral data.
Detailed features in the near-IR spectrum give rise to the difference in the best-fitting parameters for the two data sets.
Adding the $J$- and $H$-band photometry to the above data set does not improve the situation (the last row of Table \ref{sps_params});
the derived $t$, $\tau$ and $E_{\rm B-V}$ are tend to be even smaller than the previous case.
These three parameters are clearly degenerated and the different data sets find the best SED models on the different locations of this ($t$, $\tau$, $E_{\rm B-V}$)
degeneracy sequence.
On the other hand, the total stellar mass $M_{*}$ is found to be fairly robust.
It is consistent with the results of the previous studies \citep[e.g.,][]{conselice07} reporting the robustness of the stellar-mass estimate as opposed
to other parameters (galaxy age, star-formation duration, metallicity, etc.) suffering from various degeneracies.

We conclude that the optical spectrum supplemented with the near-IR broad-band data provide slightly different view of the target galaxy from 
those deduced from the whole spectral data. 
However, the basic classification of the galaxy could be achieved without the near-IR spectrum; the three
data sets listed in Table \ref{sps_params} consistently suggest that the galaxy is a few Gyr old system with faded star-formation activity and with
little dust. The stellar-mass estimate is fairly robust.
We note that the errors of the $JHK_s$ broad-band magnitudes used above have been calculated by appropriately propagating the error of the GNIRS spectrum, 
which gives 0.05 mag for the $J$ band and 0.03 mag for the $H$ and $K_s$ bands.
These are similar to the typical magnitude errors achieved in recent photometric observations.
Thus the above data sets of the optical spectrum with the near-IR broad-band photometry well represent those obtained in real observing programs.


\section{Summary\label{sec:summary}}

We present a Gemini/GMOS and GNIRS spectrum of the galaxy TSPS J1329$-$0957, a red and bright member of the ERO population.
It has been discovered in the course of the Tokyo-Stromlo Photometry Survey (TSPS) we are developing in the southern sky.
The spectrum covers red-visible to near-IR (0.6 -- 2.3 $\mu$m) wavelength range, providing a useful example to be compared with the same
population which has been vigorously discovered in the recent surveys using the 8-m class telescopes and multi-object spectrographs.
The \ion{Ca}{2} HK absorption lines are clearly detected, which gives the redshift of the galaxy as $z$ = 1.26.
The detection of \ion{Ca}{2} HK lines, the absence of any strong emission features, and the overall SED indicate that the galaxy is an old, 
passively-evolving system whose light is dominated by an evolved stellar population.
We compared the observed spectrum with the predictions of the SPS models and found that the model with
age 2 --3 Gyr, little amount of dust ($E_{\rm B-V} \la$ 0.1 mag), and  stellar mass $M_{*}$ = $10^{11.5} M_{\odot}$ to best-fit the observations.
The inferred properties of the galaxy are supported by the estimate from the  D4000 index and by the resemblance of the observed spectrum to several
composite spectra of  early-type galaxies at $z <$ 1.
We also investigated the effectiveness of using near-IR broad-band data, instead of the spectral data, in deriving the galaxy properties.
It shows that the optical spectrum supplemented with near-IR broad-band data provide slightly different view of the target galaxy from those deduced 
from the whole spectral data. 
However, the basic classification of the galaxy could be achieved without the near-IR spectrum and the stellar-mass estimate is fairly robust.

We conclude that TSPS J1329$-$0957 is very likely to have formed 2 -- 3 Gyr before it was observed, and subsequently evolved passively 
to become one of the most massive objects, of early-type, in the $z$ = 1 - 2 universe.
The presence of such a massive galaxy in the early universe might argue against the current representations of the hierarchical formation scenario and 
might favor the monolithic scenario.



\acknowledgments
We are grateful to the referee for giving useful and suggestive comments.
We thank the staff of the Siding Spring Observatory, the South African Astronomical Observatory, and
the Gemini Observatory for technical support and assistance with the observations.
The IRSF/SIRIUS team in Nagoya University, Kyoto University, and National Astronomical Observatory of Japan
has provided great help during the IRSF observations.
We are especially grateful to T. Nagata for his cooperation.
We also thank C. Koen for conducting the follow-up observation of our target.
Y. Hamada has contributed greatly to the TSPS project.
Y.M. acknowledges grants-in-aid from the Research Fellowships of the Japan Society for the Promotion of Science
(JSPS) for Young Scientists.
This work has been supported in part by Grants-in-Aid for Scientific Research
(15253002, 17104002) and the Japan-Australia Research Cooperative Program from JSPS.






\clearpage



\clearpage

\begin{table}
\begin{center}
\caption{Best-Fitting SPS Model Parameters\label{sps_params}}
\begin{tabular}{lcccc}
\tableline\tableline
Data Set & Age $t$ [Gyr] & $\tau$ [Gyr] & $E_{\rm B-V}$ [mag] & log $M_*$ [$M_{\odot}$]\\
\tableline
Optical spec. + NIR spec.   & $2.4^{+0.2}_{-0.5}$ & $0.55^{+0.10}_{-0.15}$ & $0.06^{+0.07}_{-0.03}$ & $11.49^{+0.01}_{-0.04}$\\
Optical spec. + $K_s$ phot.   & $1.6^{+0.4}_{-0.3}$ & $0.01^{+0.31}_{-0.01}$ & $0.00^{+0.02}_{-0.00}$ & $11.46^{+0.02}_{-0.04}$\\
Optical spec. + $JHK_s$ phot. & $1.3^{+0.1}_{-0.1}$ & $0.01^{+0.02}_{-0.01}$ & $0.00^{+0.02}_{-0.00}$ & $11.32^{+0.02}_{-0.01}$\\
\tableline
\end{tabular}
\end{center}
\end{table}

\end{document}